# Classification of Histopathology Images of Lung Cancer Using Convolutional Neural Network (CNN)

Neha Baranwal, Preethi Doravari and Renu Kachhoria

https://orcid.org/0000-0002-1113-7884

## Abstract

Cancer is the uncontrollable cell division of abnormal cells inside the human body, which can spread to other body organs. It is one of the non-communicable diseases (NCDs) and NCDs accounts for 71% of total deaths worldwide whereas lung cancer is the second most diagnosed cancer after female breast cancer. Cancer survival rate of lung cancer is only 19%. There are various methods for the diagnosis of lung cancer, such as X-ray, CT scan, PET-CT scan, bronchoscopy and biopsy. However, to know the subtype of lung cancer based on the tissue type H and E staining is widely used, where the staining is done on the tissue aspirated from a biopsy. Studies have reported that the type of histology is associated with prognosis and treatment in lung cancer. Therefore, early and accurate detection of lung cancer histology is an urgent need and as its treatment is dependent on the type of histology, molecular profile and stage of the disease, it is most essential to analyse the histopathology images of lung cancer. Hence, to speed up the vital process of diagnosis of lung cancer and reduce the burden on pathologists, Deep learning techniques are used. These techniques have shown improved efficacy in the analysis of histopathology slides of cancer. Several studies reported the importance of convolution neural networks (CNN) in the classification of histopathological pictures of various cancer types such as brain, skin, breast, lung, colorectal cancer. In this study tri-category classification of lung cancer images (normal, adenocarcinoma and squamous cell carcinoma) are carried out by using ResNet 50, VGG-19, Inception_ResNet_V2 and DenseNet for the feature extraction and triplet loss to guide the CNN such that it increases inter-cluster distance and reduces intra-cluster distance.
**Keywords:** ResNet 50, CNN, VGG-19, Inception_ResNet_V2 and DenseNet, Histopathology Images

## Introduction

Cancer is the uncontrollable cell division of abnormal cells inside the human body, which can spread to other body organs. The process of transformation of normal cells into cancerous cells due to genetic alteration is known as Carcinogenesis as shown in Figure 1. The process of carcinogenesis occurs in three phases. The first is the Initiation phase, where any alterations that occur in the normal cell due to gene mutation can cause a change in gene expression and even deletion of a part of Deoxyribonucleic acid (DNA) sometimes. If these changes skip the repair mechanism during the cell cycle, then the cell with altered genes remains as it is. In the Promotion phase, which is the second phase, the altered cell starts proliferation. In the final stage, the Progressive phase the cells start proliferating aggressively by number, size, and form primary tumors. In this stage, the cells become invasive and metastatic. Phases of carcinogenesis is shown in Figure 2 (Chegg.com, 2021).

<Figure 1 here>

The name for a cancer type is given based on the body organ or the cell type from which cancer originates. So far, there are more than 100 types of cancer found. There are various types of cancer such as breast, brain, lung, colon cancer, etc. Cancer is one of the non-communicable diseases (NCDs) and NCDs account for 71% of total deaths worldwide (World Health Organization, 2019). Whereas lung cancer is the second most diagnosed cancer after female breast cancer (Ferlay, 2021). According to GLOBOCAN 2018, 2.09 million new lung cancer cases have been reported and accounted for 1.76 million deaths globally, resulting in the highest mortality rate in both males and females when compared to other cancer types (Bray, 2018). The incidence of lung cancer is higher among young women when compared to young men in the United States (Jemal, 2018). Approximately 63,000 lung cancer cases are recorded each year in India (Noronha, 2012). The cancer survival rate of lung cancer is only 19% (Siegel, Miller, and Jemal, 2019).

<Figure 2 here>
 Lung cancer is divided into two major types based on histology, biological behaviour, prognosis and treatment. They are non-small cell lung cancer (NSCLC) and Small cell lung cancer (SCLC). NSCLC is the most common cancer type, which accounts for 85% and the remaining 15% is SCLC. NSCLC is again sub-divided into adenocarcinoma, squamous cell carcinoma and large cell carcinoma. As shown in Figure 3, adenocarcinoma is the most common cancer type and it is formed in epithelial cells that secrete mucus or fluids. Whereas in squamous cell carcinoma, cancer originates from squamous cells that line many organs such as the lung, bladder, kidney, intestines, and stomach (Pêgo-Fernandes, 2021; Cancer.gov, 2007).

<Figure 3 here>
There are various methods for the diagnosis of lung cancer, such as X-ray, CT scan, PET-CT scan, bronchoscopy and biopsy. However, to know the subtype of lung cancer based on the tissue type H and E staining is widely used, where the staining is done on the tissue aspirated from a biopsy. Hematoxylin (H) has a deep purple colour, stains nucleic acids in the cells and Eosin (E) have pink colour, and it stains proteins. (Fischer et al, 2008). Studies have reported that the type of histology is associated with prognosis and treatment in lung cancer (Hirsch et al, 2008; Itaya et al, 2007; Weiss et al, 2007). Recent advances in genomic studies paved the path to personalized medicine for lung cancer patients (Travis et al, 2021; Galli and Rossi, 2020).

Therefore, early and accurate detection of lung cancer histology is an urgent need and as its treatment is dependent on the type of histology, molecular profile and stage of the disease, it is most essential to analyse the histopathology images of lung cancer. However, manual analysis of histopathology reports is time-consuming and subjective. With the advent of personalized medicine, pathologists are finding it difficult to manage the workload of dealing with a histopathologic cancer diagnosis. Hence, to speed up the vital process of diagnosis of lung cancer and reduce the burden on pathologists, Deep learning techniques (Baranwal et al, 2019, Tripathi et al, 2013, Kumud et al., 2015 and singh et al, 2020) are used. These techniques have shown improved efficacy in the analysis of histopathology slides of cancer (Litjens et al, 2016).

**Analysis of Previous Research**

In the next few decades, cancer is expected to be the leading cause of death and is one of the biggest threats to human life (Tang et al, 2009). To improve the efficiency and speed of cancer diagnostics, Computer-aided diagnosis (CAD) was applied to the analysis of clinical data. There has been vast development in the field of CAD and many machine learning techniques are developed for the diagnosis purpose. Among all machine learning techniques, neural networks have shown increased performance in the detection of medical images. In the classification of lung cancer images, different CNN algorithms are used to improve the accuracy of the prediction and classification. Such accurate predictions aid doctors by reducing the workload and prevent human errors in the process of diagnosis.

a) **Computer aided diagnosis in medicine:** Computer-aided diagnosis (CAD) is cutting-edge technology in the field of medicine that interfaces computer science and medicine. CAD systems imitate the skilled human expert to make diagnostic decisions with the help of diagnostic rules. The performance of CAD systems can improve over time and advanced CAD can infer new knowledge by analysing the clinical data. To learn such capability the system must have a feedback mechanism where the learning happens by successes and failures. During the last century, there is a dramatic improvement in human expertise and examination tools such as X-ray, MRI, CT, and ultrasound. With the discovery and study of new diseases and their progression, the diagnosis has become difficult and more complex. Various factors such as complex medical diagnosis, availability of vast data pertinent to conditions and diseases in the field of medicine, increasing knowledge on diagnostic rules, and the emergence of new areas such as AI, machine learning, and data mining in the field of computer science has led to the development of CAD (Yanase and Triantaphyllou, 2019a). Quantitative analysis of pathology images has gained importance among researchers in the field of pathology and image analysis. There is clearly a need for quantitative image-based evaluation of pathological slides as the diagnosis is based on the opinion of pathologists. CAD can reduce the burden on pathologists by filtering out the benign cancer images so that the pathologists can focus on more complicated images that are difficult to diagnose and suspicious. Quantitative analysis of pathology images not only helps in diagnosis but also in medical research (Gurcan and Boucheron, 2019). At many hospitals in the United States CAD has become a part of routine clinical work for screening mammograms for the detection of breast cancer (Freer and Ulissey, 2001; Doi, 2007). In the fields of radiology and medical imaging, CAD has become the major research subject (Doi, 2007). These are cost-effective and can be used for the early detection of disease. Diseases like cancer are very aggressive when detected at later or advanced stages, hence screening and detection of such disease can avoid unnecessary invasive procedures for the treatment of the disease. Moreover, these models can eliminate human errors such as the detection of microcalcifications and help to improve the workflow of diagnostic screening procedures (Nishikawa et al, 2012; Yanase and Triantaphyllou, 2019a).

b) **CNN and cancer image detection:** In the field of medicine to improve the quality of patient care machine learning-based approaches are used. These approaches are used to analyze and evaluate the complex data. The applications of artificial intelligence can speed up support delivery, be cost-effective, and at the same time can reduce medical errors (Jia et al, 2016). Recent studies revealed that advances in Artificial Intelligence (AI) have exceeded human performance in various fields and domains (Fogel and Kvedar, 2018) such as human robot interaction (Baranwal et al, 2017 and Singh et al, 2021), face recognition(Baranwal et al, 2019, Baranwal et al, 2016 and Singh et al, 2017) etc. Several studies reported the importance of convolution neural networks in the classification of histopathological pictures of various cancer types such as brain, skin, breast, lung, colorectal cancer (Garg and Garg, 2021; Mobadersany et al, 2018). Convolution neural networks have exceeded even human performance on ImageNet Large-Scale Visual Recognition Challenge (ILSVRC) and performed well in classification with second best error rate (Lundervold and Lundervold, 2019). Deep convolutional neural network (DCNN) models for the classification of lung cancer images showed increased accuracy and reduced model overfitting by using various data augmentation techniques (Teramoto et al, 2017). A study that used LC25000 and Colorectal Adenocarcinoma Gland (CRAG) datasets to train and classify the histopathology images reported the highest sensitivity using ResNet-50 (96.77%) which is followed by ResNet-30 and ResNet-18 with 95.74% and 94.79% sensitivity respectively (Bukhari et al, 2020). Another study used low dose computed tomography (LDCT) images for the detection of early lung cancer, where they reported 97.5 % of sensitivity where the SVM model was used for classification where VGG19 was used for feature extraction. As the image dataset small, they used transfer-learning methods to obtain better prediction results (Elnakib, Amer, and Abou-Chadi, 2020). Satvik Garg and Somya Garg developed eight Pre-trained CNN models that include various feature extraction tools such as VGG16, InceptionV3, ResNet50, etc. for the classification of lung and colon cancer images and achieved accuracies ranging from 96% to 100%. To boost the performance of the model better augmentation technique, an imaging library was used (Garg and Garg, 2021). Homology-based image processing (HI) model for the multicategory classification of lung cancer images achieved better accuracy when compared to conventional texture analysis (TA). For feature extraction in the HI model, Betti numbers are the important metrics (Nishio et al, 2021). A convolution neural network model with cross-entropy as a loss function achieved training and validation accuracy of 96.11% and 97.2% for the classification of lung cancer images (Bijaya Kumar Hatuwal and H.C.T, 2021). The combination of Deep Learning and Digital Image Processing for the classification of lung and colon cancer histopathology images obtained maximum accuracy of 96.33% (Masud et al, 2021). In the classification of histopathological images of colorectal cancer ResNet-50 along with transfer, learning reported an accuracy of 97.7% which is so far the best when compared to all previous results in the literature (Shawesh and Chen, 2021). In the classification of histopathology images of breast cancer, Inception_ResNet_V2 has proved to be the best deep learning architecture (Xie et al, 2019).

| Cancer type | Contribution | Technique used |
|---|---|---|
| Leukemia, Colon cancer | Gene Selection for Cancer Classification | SVM technique based on Recursive Feature Elimination (RFE) (Guyon et al, 2002) |

| Lymphoma Data, SRBCT, Liver Cancer, Different tumor types | Finding the smallest set of genes | Gene Importance Ranking, Support Vector Machines (SVMs) (Wang, Chu and Xie, 2007) |
|---|---|---|
| Leukemia, Colon, and Lymphoma | Cancer classification | Ensemble of neural networks (Cho and Won, 2007) |
| Ovarian cancer | Ovarian cancer diagnosis | Fuzzy neural network (Tan, Quek, Ng and Razvi, 2008) |
| Prostate cancer, lymphoma, Breast cancer | Gene Prioritization and Sample Classification | Rule-Based Machine Learning (Glaab, Bacardit, Garibaldi and Krasnogor, 2012) |
| Microarray data of six cancer types (leukemia, lymphoma, prostate, colon, breast, CNS embryonal tumor) | Gene selection and classification | Recursive Feature Addition, Supervised learning (Liu et al, 2011) |
| Microarray data of multiple cancer types | Cancer classification | Particle swarm optimization, Decision tree classifier (Chen, Wang, Wang and Angelia, 2014) |
| Multiple cancer types | Cancer Classification | Ensemble-based Classifiers (Margoosian and Abouei, 2013) |
| breast cancer | Cancer Classification | Deep Belief Networks (Zaher and Eldeib, 2016) |
| Leukemia | Cancer classification | Artificial neural network (ANN) (Dwivedi, 2018) |
| Gene expression data from multiple cancer types. | Molecular Cancer Classification | Transfer Learning, Deep Neural Networks (Sevakula et al, 2018) |
| Breast Cancer | Breast Cancer Classification | Convolutional Neural Network (Ting, Tan and Sim, 2019) |

| Cervical cancer | Cervical cancer classification | Convolutional neural networks & extreme learning machines (Ghoneim, Muhammad and Hossain, 2020) |
| --- | --- | --- |
| Melanoma | Automated Melanoma Recognition | Deep Residual Networks (Yu et al, 2016) |
| Breast Cancer | Breast Cancer Detection | Deep Learning From Crowds for Mitosis (Albarqouni et al, 2016) |
| Cervical cancer | Classification of cervical Pap smear images | Mean-Shift clustering algorithm and mathematical morphology (Wang et al, 2019) |
| Cervical cancer | Cervical Cell Classification | Deep Convolutional Networks (Zhang et al, 2017) |

Table 1. Summary of classification of various cancer types using machine learning techniques (Sharma and Rani, 2021).

Hence there is a need to explore different techniques to improve the model performance other than increasing parameters. In the classification of images of cancer, there should be immense effort to differentiate cancer images from non-cancer images. The accuracy of the model needs to be high in such cases and the model should be able to detect both intra-class diversity and inter-class similarity. To consider such factors and guide the model accordingly, FaceNet introduced triplet loss (Schroff, Kalenichenko, and Philbin, 2015).

**Proposed Research Work**

To classify the lung cancer images, the dataset is obtained from LC25000 Lung and colon histopathological image dataset which is already augmented data having 5000 images in each class of lung cancer image set comprising three classes. This dataset is pre-processed using python tools and features are extracted by CNN techniques, later the model is created and evaluated. Various CNN techniques are used to compare and classify the images. Complete flow of proposed method is shown in Figure 4.

<Figure 4 here>

**a) Dataset description:**

Data is drawn from the LC25000 Lung and colon histopathological image dataset, which consists of 5000 images each in three classes of benign (normal cells), adenocarcinoma and squamous carcinoma cells (both are cancerous cells). The dataset is HIPAA compliant and validated (Borkowski et al, 2019). The original images obtained are only 750 images in total and the size of the images are 1024 x 768 pixels, where each category gets 250 each. These images are cropped to 768 x 768 pixels using python and expanded using the augmentor software package. Thus, the expanded dataset contains 5000 images in each category. Augmentation is done by horizontal and vertical flips and by the left and right rotations (Borkowski et al, 2019). The sample images for each category are shown in Figure 5.

<Figure 5 here>

b) **Data Pre-processing:**

Data pre-processing is an essential step, which helps in improving the quality of the images and it includes data preparation, data normalization, data cleaning, and data formatting. Data preparation aids in the transformation of data by modifying it into the appropriate format. Whereas data normalization makes a different image format into a regular format where all the images are uniform while in data transformation, the data is compressed (Zubi and Saad, 2011). As the images are already augmented, ImageDataGenerator which is imported from Keras. Preprocessing, image class used for the preprocessing of the image dataset. A total of 15000 images are used for the train-test split, in which 80% of the images are used for training and 20% for validating the data.

c) **Feature extraction:**

Feature extraction is used to decrease the model complexity where important features are recognized from the images. For the knowledge extraction from images, not all the features provide interesting rules for the problem. This is the major step where the model performance and effectiveness are dependent. To extract such features as color, texture, and structure, image-processing techniques are used. This can be achieved by localizing the extraction to small regions and ensuring to capture all areas of the image (Zubi and Saad, 2011). For feature extraction, ResNet 50(He et al, 2016), VGG19(Munir et al, 2019), Inception_ResNet_V2 (Xie et al, 2019; Kensert, Harrison and Spjuth, 2019), DenseNet121(Huang, Liu, Van Der Maaten, and Weinberger, 2017; Chen, Zhao, Liu and Lin, 2021) is used.

D) **Loss function:**

For a machine learning model to fit better while training the neural networks, loss function acts as a major key for adjusting the weights of the network. During the back propagation while training, loss function penalizes the model if there is any deviation between the label predicted by model and the actual target label (https://ieeexplore.ieee.org/abstract/document/8943952). Hence the use of loss function is very critical to achieve better model performance. Triplet loss is used as loss function in this study.

**Triplet loss:**

Triplet loss is first developed for face recognition by Schroff et al, 2015 by mapping Euclidean distance to find the similarities in the face images. Although the images are blurred with the help of the distances between faces of similar and different identities this method can be used (Schroff, Kalenichenko and Philbin, 2015). To increase the inter-cluster similarity and intra-cluster diversity, triplet loss is used as a cost function to guide the learning of Convolutional neural networks. It can increase the inter-class distance and decrease the intra-class aiding the classification process of the model. In equation 1, a and p are the vectors that belong to the same category, whereas n is a vector that belongs to another category.

$$L_t = (d(a,p) - d(a,n) + margin, 0) \tag{1}$$

From the above formula, we can say that the triplet loss guides the model to shorten the distance between images of the same category and increases the distance between images that belong to different categories (Zhang et al, 2020). It has been reported that the use of triplet loss shown improved accuracy in binary classification when compared to using the base model (Agarwal, Balasubramanian and Jawahar, 2018).

### e) Model and evaluation metrics

A CNN is created using a stack of layers for image recognition and classification. Before passing through the fully connected layer, the training and testing data is passed through parameters such as max-pooling and kernel filters. Activation function ReLU is used in all three hidden layers and a softmax function is applied to classify the images.

In order to evaluate the performance of the model the following metrics are measured:

Accuracy: Over the total number of data instances accuracy represents the correctly classified data. Equation (2) represents the formula to calculate accuracy. However, accuracy alone may not be a good measure to decide the performance of the model.

Precision: This is used to measure the positive predictive observations. It represents the correctly predicted positive observations of total predicted positive observations. Equation (3) is the formula to calculate the precision. High precision relates to a low false-positive rate.

Recall (Sensitivity): Recall represents the correctly predicted positive observations of total actual positive observations. The formula to calculate recall is given in Equation (4). It is also known as sensitivity or true positive rate.

F1 score: Ideally, a good evaluation should consider both precisions and recall to seek balance. A weighted average of precision and recall is the F1 score. Equations (5) is the formula to calculate the F1 score. For uneven class distribution, the F1 score is more useful to evaluate the model.

$$\text{Accuracy} = (TP + TN)/((TP + FP + FN + TN)) \tag{2}$$

$$\text{Precision} = TP/((TP + FP)) \tag{3}$$

$$\text{Recall} = TP/((TP + FN)) \tag{4}$$

$$\text{F1 Score} = (2*(Recall * Precision))/((Recall + Precision)) \tag{5}$$

### Result and analysis

All four CNN architecture models have been trained using specific and fine-tuned parameters to achieve better model performance. Initially pre-trained CNN architecture is used to classify the lung cancer cells. In these models' cross entropy is used as loss function. VGG19 model is trained by adding two hidden layers with embeddings 256 and 128 with ReLU as an activation function and for the final output layer softmax is used as activation function. For this model cross entropy is used to calculate the loss over 18 batch size. When the model is trained with 30 epochs with Adam as an optimizer (in default setting), it showed validation loss of 0.196. The performance of the model has shown accuracy of 92.1%, precision of 92.5%, recall of 92.1% and f1 score of 92.04% on validation dataset. Similarly, ResNet50 model is trained using the same number of hidden layers as VGG19. All the parameters are same for both the models and when the model is trained for 30 epochs the validation loss showed by the model is 0.03. Among all ResNet has shown improved performance when compared to VGG19 model. This model showed accuracy, precision, recall and f1 score of 99%. Inception-ResNetv2 is trained using two layers, in which one is global average pooling and the other one is dense layer with 1024 embeddings. The activation layer used for the hidden layer is ReLU and for the output layer is softmax. When the model is trained for 30 epochs with Adam as optimizer in default, the validation loss of the model is 0.008. The performance of this model is much better than other models, where test accuracy, precision, recall and f1score is 99.7%. Lastly, DenseNet121 model which is trained with two hidden layers of 1024 and 500 embeddings with Adam in default setting as optimizer has shown validation loss of 0.01. After evaluation of this model on test data the accuracy, precision, recall and F1score is 99.4%. These evaluation metrics are shown in Table 2 for comparison. All the four CNN architecture Inception-ResNetv2 model has shown improved performance and classified benign tissue images from cancer images without any misclassifications. The only misclassification happened is between the subclasses of lung cancer images as shown in Figure 6. Even validation loss is also very minimum for this model as shown in Figure 7.

| Evaluation metrics | VGG19 | ResNet50 | Inception-ResNetv2 | DenseNet121 |
|---|---|---|---|---|
| Accuracy | 92.1% | 99 | 99.7 | 99.4 |
| Specificity | 92.5% | 99 | 99.7 | 99.4 |
| Recall | 92.1% | 99 | 99.7 | 99.4 |
| F1 score | 92.4% | 99 | 99.7 | 99.4 |

Table 2. Evaluation metrics for all the four CNN architectures

<Figure 6 here>

<Figure 7 here>

To compare the pre-trained model with triplet neural network, again the four CNN architectures are trained using triplet neural network. In these models after train, test split, the data is divided into three images where first is the anchor, second is positive image which has same class label as anchor and the third is the negative image where the class label of this is different from anchor. For such triplet selection the loss function is introduced such that the distance between anchor and positive image should be always less than the distance between anchor and negative image. For such triplet loss function margin/alpha is added to calculate the distance. In these models this margin is set to 0.4 as while analysis, the model did not perform better at higher or lower margin other than 0.4. The batch size of input image is set to 16 and data type of each input is changed to float16 because of GPU memory constraints. After training the four models with introducing triplet selected, the learning rate of the Adam is also finely tuned to fit the model as shown in Table 5.2. For all the models Global Average pooling layer and L2 Normalization is used.

| Model | Adam-Learning rate used |
|---|---|
| VGG19 | 0.00001 |
| ResNet50 | 0.0001 |
| Inception-ResNetv1 | 0.00001 |
| DenseNet121 | 0.0001 |

Table 3: Fine tuning of learning rate for different CNN models

All four models are trained for 10 epochs using 150 steps in each epoch and validation steps of 50. Validation loss of all the four models is mentioned in the Figure 8.

Evaluation of triplet model is done by using KNN approach, where the model embeddings from training dataset are taken and trained using Nearest Neighbors. Later the nearest neighbor for test data embeddings are predicted using the trained model. Using this class label of the predicted test data is considered for evaluating the model.

<Figure 8 here>

It is observed that DenseNet121 has shown least validation loss of all the four networks. After the evaluation of all models, highest accuracy is reported by DenseNet121 and the least by ResNet50. The evaluation metrics of the models are given in table 4. As shown in Figure 9 when the test data embeddings are plotted the DensNet121 model showed defined clusters when compared to other models.

| Evaluation metrics | VGG19 | ResNet50 | Inception-ResNetv2 | DenseNet121 |
|---|---|---|---|---|
| Accuracy | 97.69 | 96.2 | 97.04 | 99.08 |
| Specificity | 97.7 | 96.2 | 97.03 | 99.09 |
| Recall | 97.69 | 96.2 | 97.04 | 99.08 |
| F1 score | 97.69 | 96.1 | 97.04 | 99.08 |

Table 4: Evaluation metrics for all the four CNN architectures

<Figure 9 here>

**Conclusion**

CNN models have shown to increase accuracy with fine tuning of hyper parameters. Various CNN architectures are compared in the study to get better accuracy and to compare which architecture gives better performance for this dataset. Model performance of all four CNN models such as VGG19, ResNet50, Inception-ResNetv2 and DenseNet121 have shown increased accuracy. Although the pre-trained models are available, fine-tuning of these models are necessary to obtain desired results. In this study Inception-ResNetv2 has shown a very high-test accuracy rate of 99.7% when compared to other models where the accuracy of VGG19, ResNet50 and DenseNet121 are 92,99 and 99.4% respectively. When the triplet neural network model is trained on these four pre-trained models DenseNet121 achieved test accuracy of 99.08% which is the highest of all other four. Test accuracies of other three models are 97.69, 96.2, 97.04% for VGG19, ResNet50 and Inception-ResNetv2 respectively. The obtained model with high accuracy has significantly classified cancer images from non-cancerous images which is a crucial step in cancer diagnosis. There were no misclassifications among cancer and non-cancer images. Only very few misclassifications happened among the two lung cancer subtypes, that is adenocarcinoma and squamous cell carcinoma. Although the image aspect ratio of image trained triplet neural networks is low, that is 128×128×3 and batch size is 16 due to GPU constraints, the triplet network model has shown better performance.

**Notes:** For more information about different applications of AI and deep learning techniques please go through the these papers [114][115][116][117][118][119][120][121][122][123][124][125][126][127][128][129][130][131][132][133][134][135][136][137][138][139][140][141][142][143][145][146]

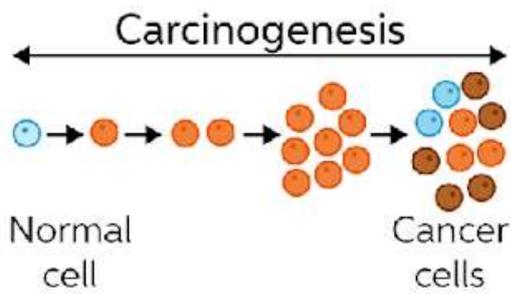

Figure 1. Process of carcinogenesis (Chegg.com, 2021).

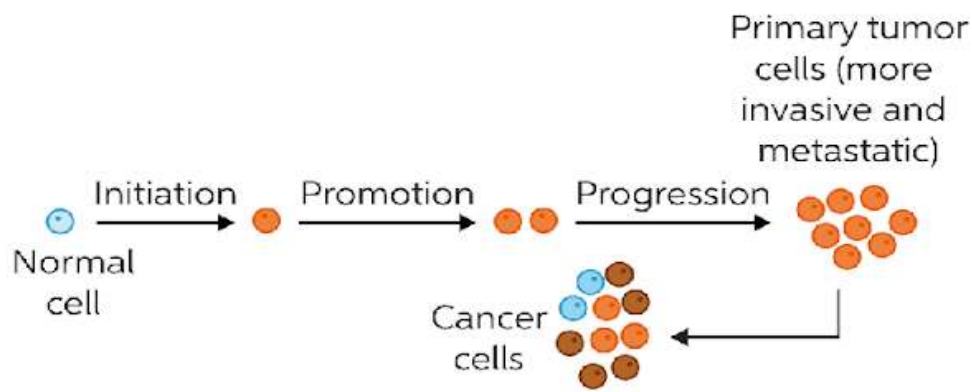

Figure 2. Three phases of carcinogenesis (Chegg.com, 2021)

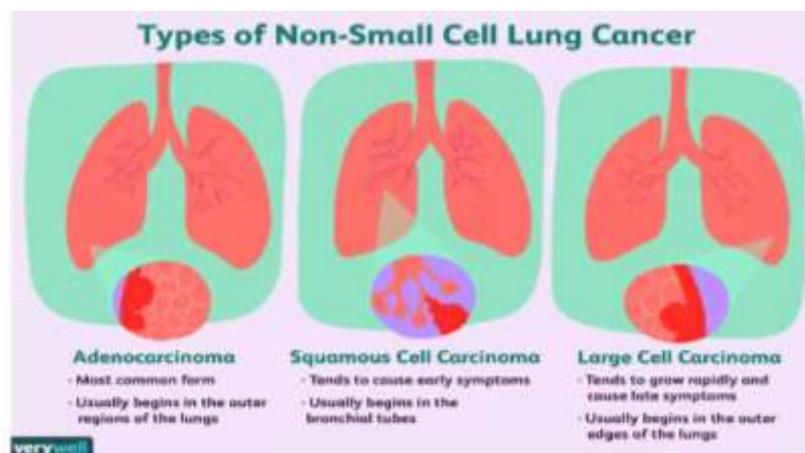

Figure 3. Types of Non-Small Cell Lung Cancer (NSCLC) (Lynne Eldridge, 2021)

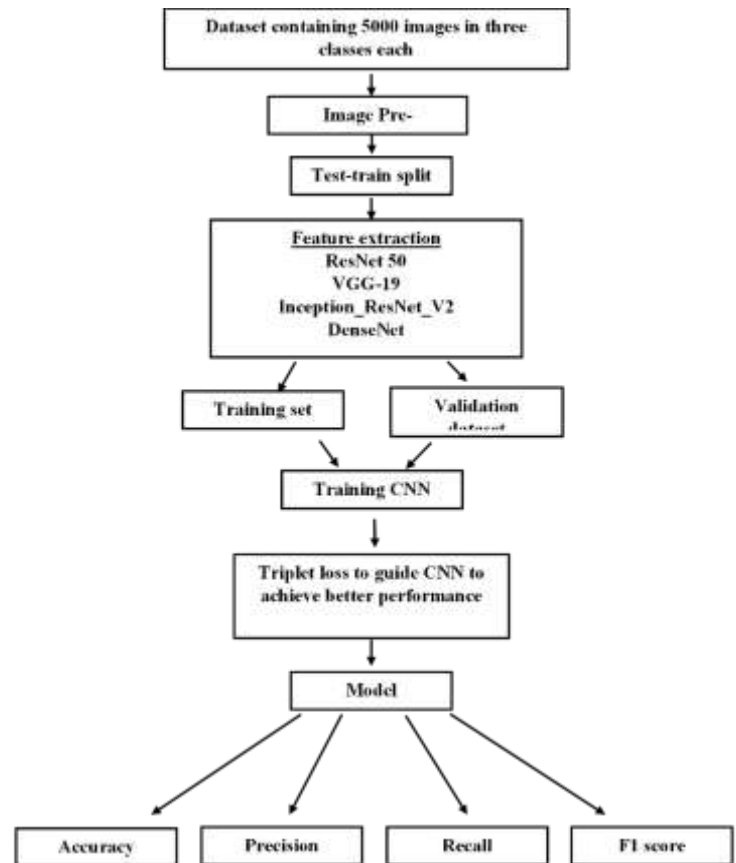

Figure 4. Proposed Methodology

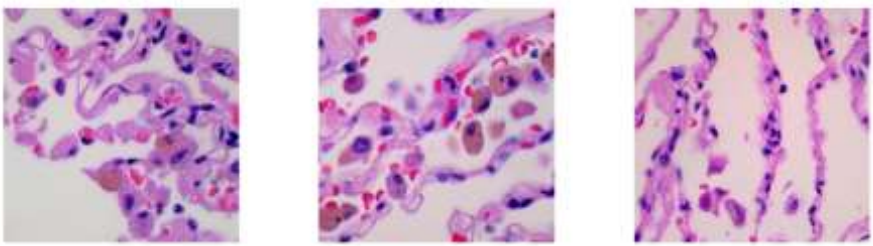

(a)

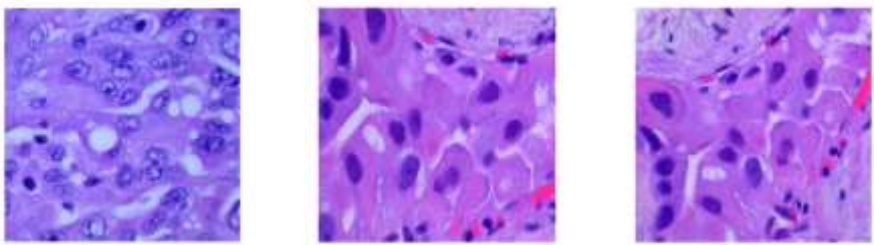

(b)

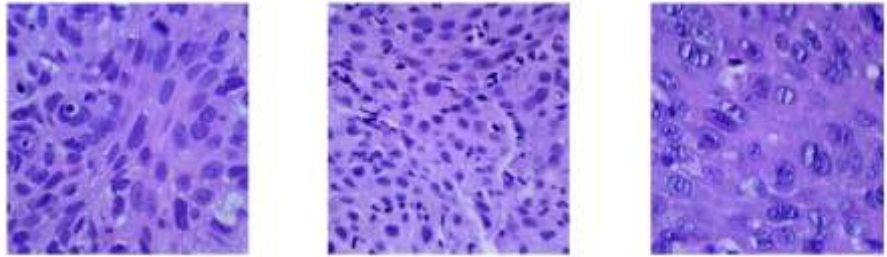

(c)

Figure 5. Sample images of three classes present in the dataset. (a) lung_n (lung normal cells),
(b) lung_aca (lung adenocarcinoma cells) and (c) lung_scc (lung squamous cell carcinoma.

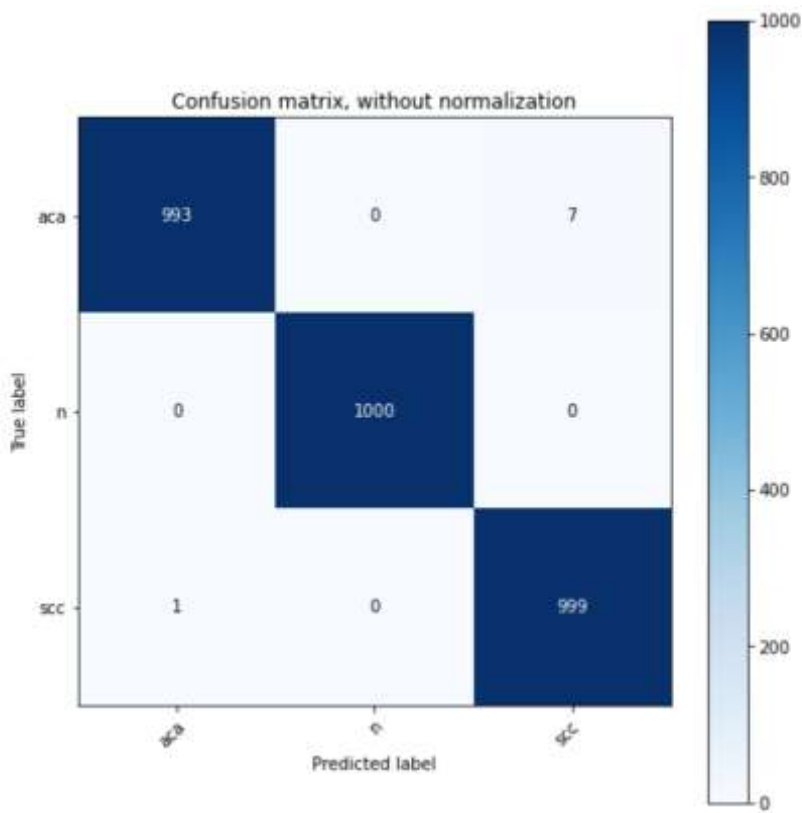

Figure 6. Confusion matrix of test data of Inception-ResNetv2ple images

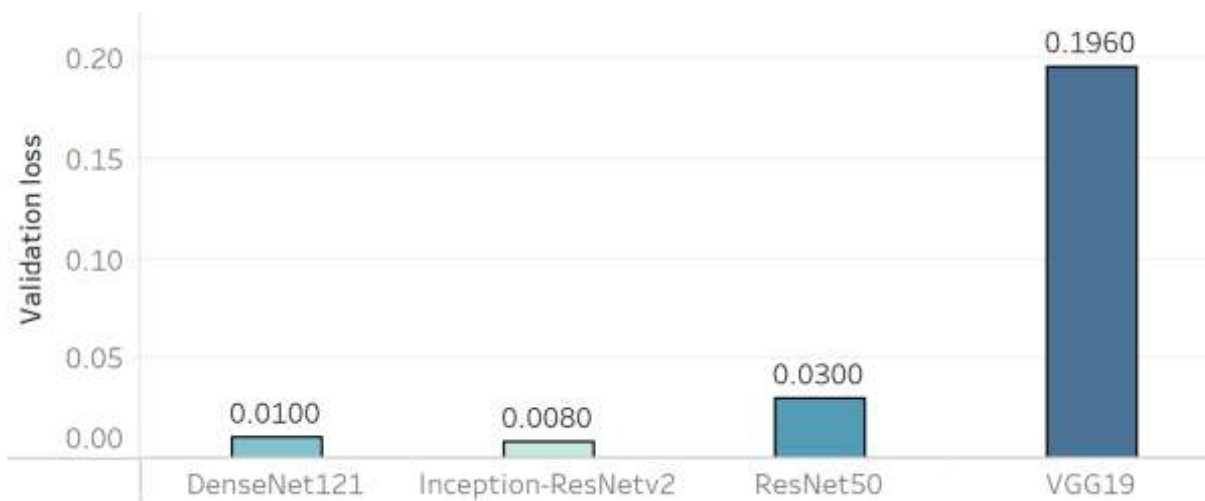

Figure 7. Validation loss obtained after training four CNN architectures

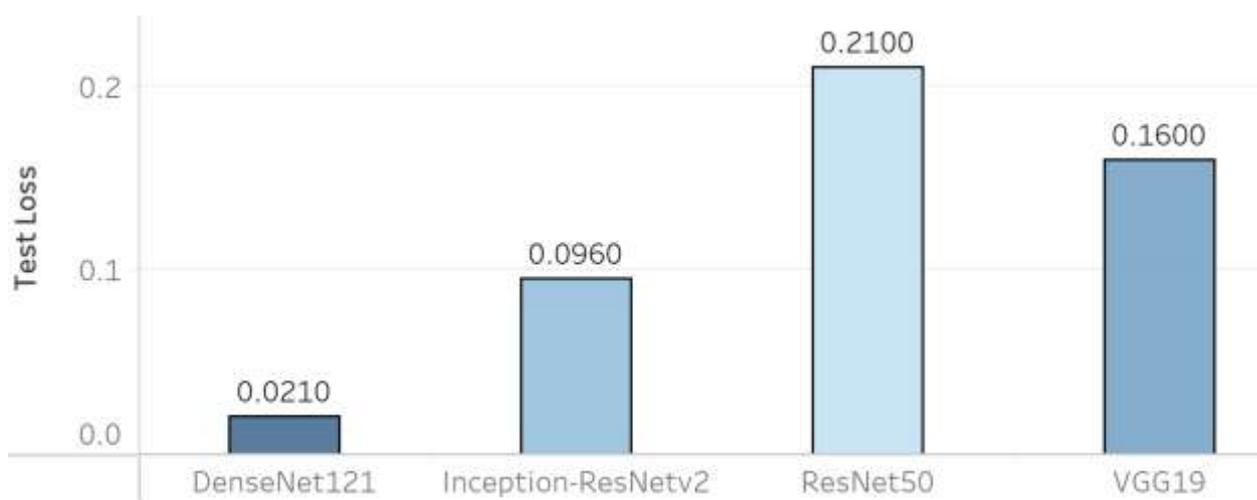

Figure 8. Validation loss of four CNN architectures trained on triplet neural network.

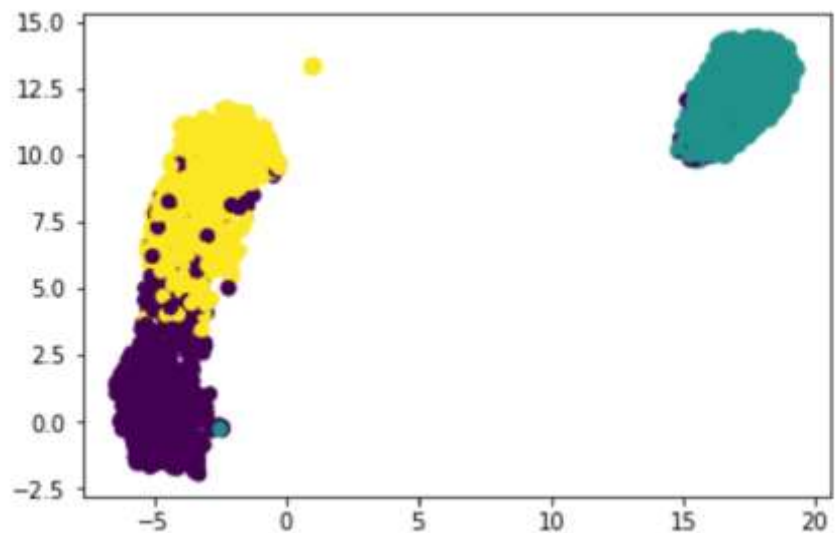

Figure 9: Clusters obtained when test embeddings are plotted. (2D array of embeddings are plotted along x and y axis)